\documentclass[aps,prd,floatfix,nofootinbib,superscriptaddress,showpacs,twocolumn]{revtex4-1}

\usepackage{amssymb}
\usepackage[intlimits]{amsmath}
\usepackage{amsfonts}
\usepackage{dsfont}
\usepackage{subfigure}
\usepackage[colorlinks=true,linkcolor=blue,citecolor=blue,urlcolor=blue]{hyperref}
\usepackage{graphicx}
\usepackage[usenames]{color}
\usepackage{natbib}
\usepackage{multirow}
\usepackage{verbatim}
\usepackage{slashed}
\usepackage{nicefrac}
\usepackage[latin1]{inputenc} 

\newcommand{\conjg}[1]{\ensuremath{\hspace{1pt}\overline{\hspace{-1pt}#1\hspace{-1pt}}}\hspace{1pt}}
\newcommand{\mc}[1]{\mathcal{#1}}

\def\q{\text{q}}

\newcommand{\Tr}{\,\mathrm{Tr}\,}

\definecolor{violet}{RGB}{111,0,255}
\definecolor{webgreen}{rgb}{0,0.75,0}
\definecolor{webred}{rgb}{0.75,0,0}
\definecolor{webblue}{rgb}{0,0,0.75}
\definecolor{darkblue}{rgb}{0,0,0.6}
\definecolor{darkgreen}{rgb}{0,0.5,0.5}
\definecolor{darkpurple}{rgb}{0.5,0,0.5}
\definecolor{darkorange}{rgb}{1,0.5,0}
\definecolor{darkgrey}{rgb}{0.4,0.4,0.4}
\definecolor{lgray}{rgb}{0.95,0.95,0.95}
\definecolor{lgreen}{rgb}{0.95,1.00,0.90}
\definecolor{lred}{rgb}{1.00,0.90,0.80}
\definecolor{lblue}{rgb}{0.2,0.35,1.00}
\definecolor{shadecolor}{rgb}{1.00,0.92,0.82}

\definecolor{violet}{RGB}{111,0,255}
\definecolor{dgreen}{rgb}{0.1,0.50,0.1}

\begin{document}

\title{Kaon-box contribution to the anomalous magnetic moment of the muon}

\author{Gernot Eichmann}
\email[e-mail: ]{Gernot.Eichmann@tecnico.ulisboa.pt}
\affiliation{CFTP, Instituto Superior T\'ecnico, Universidade de Lisboa, 1049-001 Lisboa, Portugal}
\author{Christian S. Fischer}
\email[e-mail: ]{Christian.Fischer@physik.uni-giessen.de}
\affiliation{Institut f\"ur Theoretische Physik, Justus-Liebig Universit\"at Gie{\ss}en, 35392 Gie{\ss}en, Germany}
\author{Richard Williams}
\email[e-mail: ]{Richard.Williams@theo.physik.uni-giessen.de}
\affiliation{Institut f\"ur Theoretische Physik, Justus-Liebig Universit\"at Gie{\ss}en, 35392 Gie{\ss}en, Germany}

\begin{abstract}
We present results for the charged kaon-box contributions to the hadronic light-by-light (HLBL) correction of the muon's
anomalous magnetic moment. To this end we determine the kaon electromagnetic form factor within the functional
approach to QCD using Dyson-Schwinger and Bethe-Salpeter equations and evaluate the kaon-box contribution as
defined in the dispersive approach to HLBL. As an update to previous work we also re-evaluate the charged pion-box
contribution taking effects due to isospin breaking into account. Our results are
$a_\mu^{\pi^\pm-\text{box}} = -15.7 \,(2)(3)  \times 10^{-11}$ and
$a_\mu^{K^\pm-\text{box}}   = -0.48 \,(2)(4)  \times 10^{-11}$ thus confirming the large suppression of box contributions
beyond the leading pion box.
\end{abstract}

\maketitle


%
%
%
%
%
\section{Introduction}\label{sec:introduction}
    With a persistent discrepancy of about $3$--$4$ standard deviations between the theoretical Standard Model (SM)
    predictions and experimental determinations~\cite{Blum:2013xva},
    the anomalous magnetic moment $a_\mu = \frac{1}{2}(g-2)_\mu$ of the muon is a highly interesting quantity.
    In order to clarify whether this discrepancy includes contributions beyond the SM, both theory and experiment
    strive to improve accuracy and precision. Two new experiments at Fermilab~\cite{Venanzoni:2014ixa} and J-PARC~\cite{Otani:2015jra}
    aim at reducing the experimental error by a factor of four compared to the Brookhaven experiment E821~\cite{Bennett:2006fi,Roberts:2010cj}.

\begin{figure}[t]
      \begin{center}
        \includegraphics[scale = 1.2]{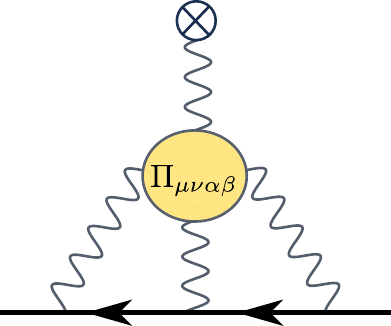}
      \end{center}
      \caption{The hadronic light-by-light scattering contribution to $a_\mu$.
      The main ingredient is the hadronic photon four-point function $\Pi_{\mu\nu\alpha\beta}$.}
      \label{fig:LBLContribution}
    \end{figure}

    In the theoretical SM calculations the error budget is dominated by QCD corrections, i.e. hadronic vacuum polarisation (HVP)
    and hadronic light-by-light (HLBL) scattering effects. The latter is shown diagrammatically in Fig.~\ref{fig:LBLContribution}. Currently
    there are great efforts both from lattice QCD~\cite{Green:2015sra,Blum:2015gfa,Green:2015mva,Asmussen:2016lse,Asmussen:2017bup,Asmussen:2018ovy,Blum:2017cer,Blum:2016lnc,Jin:2016rmu,Gerardin:2017ryf,Meyer:2018til,Gerardin:2019vio}
    as well as dispersion
    theory~\cite{Colangelo:2014dfa,Colangelo:2014pva,Colangelo:2015ama,Pauk:2014rfa,Nyffeler:2016gnb,Danilkin:2016hnh,Colangelo:2017fiz,Colangelo:2017qdm,Hoferichter:2018dmo,Hoferichter:2018kwz}
    to improve the `Glasgow-consensus' estimate of Ref.~\cite{Prades:2009tw}.

    Additional insights may be gained via the functional approach of Dyson-Schwinger and Bethe-Salpeter
    equations (DSEs and BSEs). While in all practical calculations a complete error estimate in this approach
    is very hard due to unknown systematic truncation errors, it may serve as an important cross-check
    for results in other frameworks. In addition, for contributions which cannot be accessed by the data-driven
    dispersive framework alone (e.g. due to lack of precision data), it has the potential to provide
    quantitative estimates.

    In a recent work \cite{Eichmann:2019tjk}, the functional framework has been used to determine the
    leading pseudoscalar contributions to HLBL in the dispersive framework, i.e. contributions due to
    the (on-shell) exchange of a (neutral) pion and the $\eta$ and $\eta'$ mesons. In contrast to a purely
    data-driven dispersive framework, the necessary pseudoscalar transition form factors have not been
    extracted from experiment but calculated using DSEs and BSEs. The central values of the pseudoscalar pole
    contributions to $a_\mu$ obtained in \cite{Eichmann:2019tjk} agree well within error bars with corresponding
    ones from data-driven dispersion theory \cite{Hoferichter:2018dmo,Hoferichter:2018kwz}
    and a related approach using Canterbury approximants Ref.~\cite{Masjuan:2017tvw}. In addition,
    the pion-box contribution to HLBL has been determined in the functional approach \cite{Eichmann:2019tjk}
    using the pion electromagnetic form factor calculated from the underlying quark-gluon dynamics. Again the
    result agrees with the corresponding one from data-driven dispersion theory \cite{Colangelo:2017fiz} within
    error bars.

    In this work we generalize our treatment of the meson-box contributions in two directions. First, we provide
    an updated number for the pion box, taking into account isospin breaking effects in the pion mass and, second,
    we determine the contribution from the kaon box. In model calculations using form factors from vector meson
    dominance (VMD) considerations \cite{Bijnens:1995xf} or hidden local symmetry (HLS) \cite{Hayakawa:1995ps,Hayakawa:1996ki}
    this contribution has been estimated to be suppressed by more than one order of magnitude
    as compared to the pion box. As we will see in the course of this work, this is confirmed by our approach.

    In the following we briefly summarize the technical elements of our calculation followed by a discussion
    of the results. We use a Euclidean notation throughout this work; see e.g. Appendix A of
    Ref.~\cite{Eichmann:2016yit} for conventions.

\section{Anomalous Magnetic moment}
    \begin{figure}[!b]
      \begin{center}
      \includegraphics[width=0.77\columnwidth]{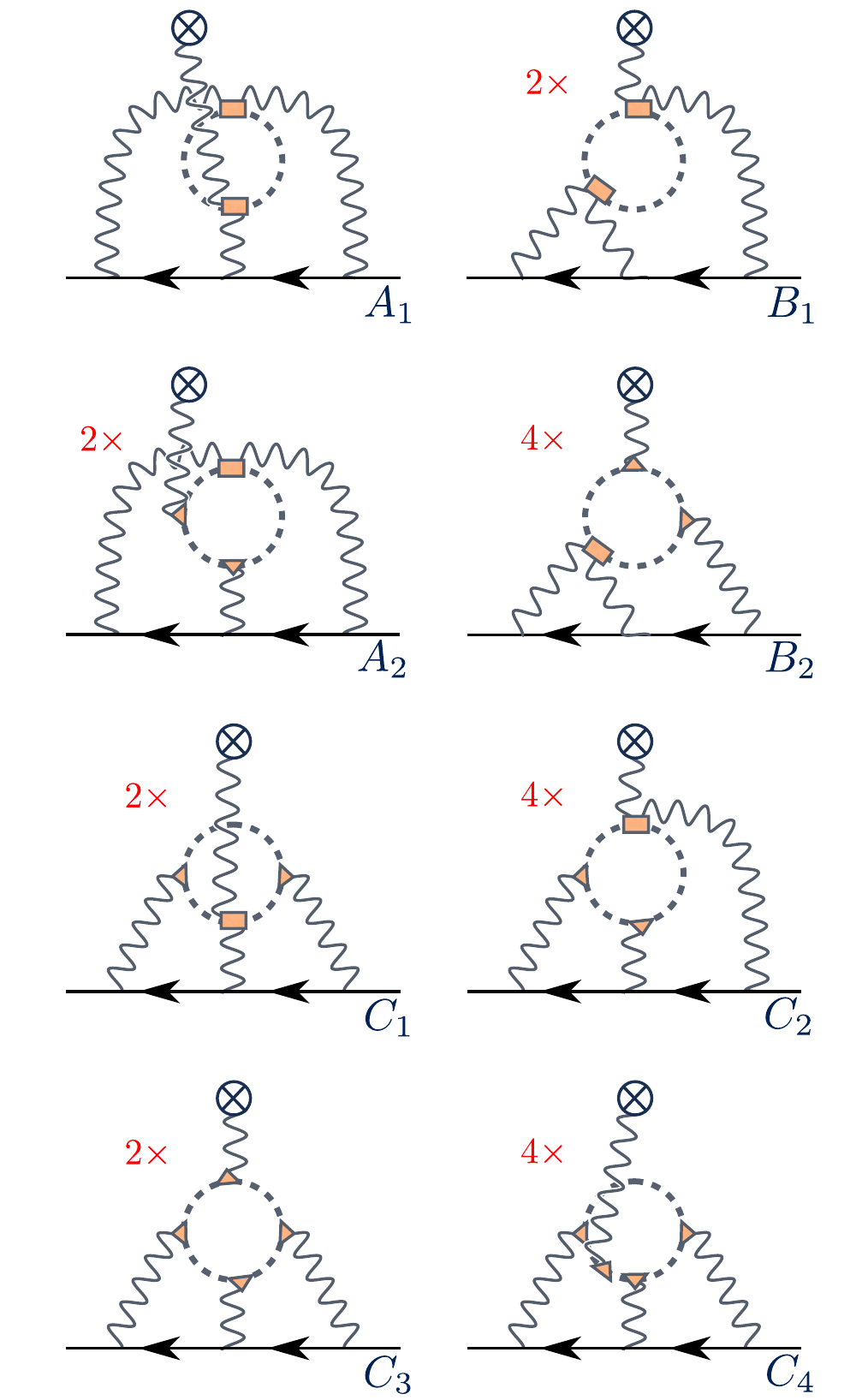}
      \caption{Meson-box contributions to the muon $g-2$ in the framework of scalar QED.\label{fig:hlbl_pion_loop} }
      \end{center}
      \end{figure}
     The anomalous magnetic moment $a_\mu$ of the muon, defined by
        \begin{align}
          a_\mu = \frac{g-2}{2}=F_2(0),
          \label{eqn:DefOfAnomaly}
        \end{align}
        is obtained from the zero momentum limit of the muon-photon vertex shown in Fig.~\ref{fig:LBLContribution}
        and decomposed on the muon mass shell according to
    \begin{align}
       \parbox{1cm}{\includegraphics[scale = 0.5]{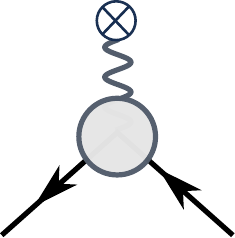}}\quad
       &=\bar{u}(p')
       \left[F_1(Q^2)\gamma^\alpha - F_2(Q^2)\,\frac{\sigma^{\alpha\beta} Q^\beta}{2 m_\mu}\right]u(p).
      \label{eqn:MuonPhotonVertexDecomposition}
    \end{align}
    Here $p$ and $p'$ are the muon momenta, $Q$ is the photon
    momentum and $\sigma^{\alpha\beta}=-\frac{i}{2}[\gamma^\alpha,\gamma^\beta]$. In order to extract $a_\mu$
    we use the technique advocated in Ref.~\cite{Aldins:1970id}, see also Ref.~\cite{Goecke:2010if,Eichmann:2019tjk}
    for details.

    Within the dispersive approach to $a_\mu$, the photon four-point vertex in Fig.~\ref{fig:LBLContribution} is decomposed
    in contributions involving one, two or more on-shell mesons coupling electromagnetically to the photon legs.
    The leading contributions in this representation
    are given by the on-shell exchange of neutral pseudoscalar mesons ($\pi_0, \eta, \eta'$) involving fully dressed
    transition form factors. These have been dealt with in a number of previous works, see e.g.
    Ref.~\cite{Masjuan:2017tvw,Hoferichter:2018dmo,Hoferichter:2018kwz,Danilkin:2019mhd, Gerardin:2019vio,Eichmann:2019tjk}.

    Here we are interested in the subleading contributions from meson-box
    diagrams. As has been demonstrated in Ref.~\cite{Colangelo:2015ama}, the meson-box topology of the dispersive
    approach coincides with the one-loop amplitude of scalar QED when coupled with meson form factors at each
    photon leg (FsQED). In the approach of \cite{Kinoshita:1984it}, which we also follow here, this requires
    the evaluation of the six classes of diagrams shown in Fig.~\ref{fig:hlbl_pion_loop} supplemented by the form factors.
    For the pion- and kaon-box contributions the only non-trivial input are thus the pion and kaon electromagnetic form factors,
    which we discuss in the next section.

    \begin{figure*}[!t]
    \begin{center}
    \includegraphics[width=0.9\textwidth]{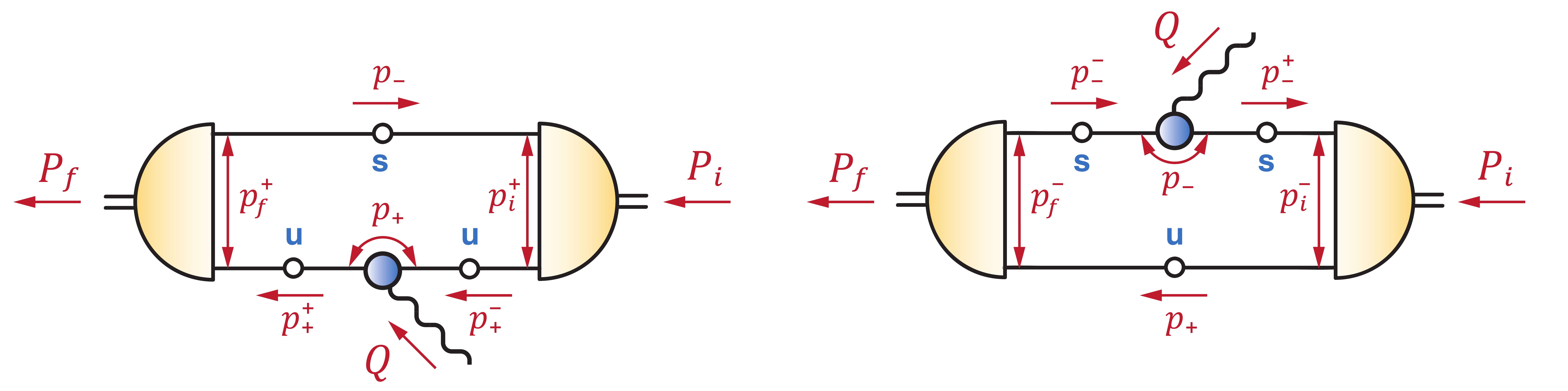}
    \caption{
    The $K^+$ electromagnetic form factor in rainbow-ladder truncation. The non-perturbative ingredients
    are the meson Bethe-Salpeter amplitude $\Gamma_M$ (yellow large half circles),
    the dressed quark propagators (straight lines) and the dressed quark-photon vertices $\Gamma^{\mu}_\q$ (blue circles).
    The internal momenta are defined in the main text.
    \label{fig.pigg:diagram} }
    \end{center}
    \end{figure*}

\section{Electromagnetic form factor of the kaon}\label{sec:meson_form_factors}

While we described the various steps needed to calculate the electromagnetic form factor of the pion in \cite{Eichmann:2019tjk},
here we detail the changes that arise for the kaon electromagnetic form factor (EMFF) in the functional DSE approach.
Further details can be found in
\cite{Maris:1999bh,Maris:2002mz,Goecke:2010if,Goecke:2012qm,Raya:2015gva,Raya:2016yuj,Eichmann:2017wil}
and the review articles \cite{Maris:2003vk,Maris:2005tt,Eichmann:2016yit}. Diagrammatically, these
form factors are calculated as shown in Fig.~\ref{fig.pigg:diagram}.

The kaon EMFF $F_K(Q^2)$ is extracted from the on-shell $\gamma K K$ current
in Fig.~\ref{fig.pigg:diagram} via
\begin{equation}\label{onshell-current}
\begin{split}
	J^\mu(P,Q) &= 2 P^\mu F_K(Q^2) \\
               &= \Tr\!\!\int_p  S_l(p_+^+)\,q_l \Gamma^{\mu}_l(p_+,Q) \,S_l(p_+^-) \\
               & \,\,\,\,\,\,\,\;\; \times \Gamma_{K}(p_i^+,P_i) \,S_s(p_-)\, \bar{\Gamma}_{K}(p_f^+,P_f)\\
               &+ \Tr\!\!\int_k  \bar{\Gamma}_{K}(p_f^-,P_f) S_l(p_+)\,\Gamma_{K}(p_i^-,P_i) \\
               & \,\,\,\,\,\,\,\;\; \times \,S_s(p_-^+)\,q_s \Gamma^{\mu}_s(p_-,Q) \,S_s(p_-^-)\, \,,
\end{split}
\end{equation}
where the index $l$ stands for a light quark $l \in {u,d}$ and $s$ for a strange quark. Furthermore,
$Q$ is the photon momentum, $P_{f,i} = P \pm Q/2$ are the final and initial kaon momenta, the relative momenta are
$p_i^\pm = p + \frac{\epsilon \mp 1}{4} Q$, $p_f^\pm = p - \frac{\epsilon \mp 1}{4} Q$ and
the quark momenta are $p_\pm = p + \frac{\epsilon \pm 1}{2} P$ and $p_\omega^\pm = p_\omega \pm Q/2$ with $\omega=\pm$.
The momentum partitioning parameter $\epsilon$ can take values between $-1$ and $1$. The electromagnetic quark charges are denoted by $q_{l,s}$. Also the quark propagators $S_{l,s}$
correspond to light or strange (anti-)quarks. Charge conjugation of the Bethe-Salpeter amplitudes is defined
by $\conjg{\Gamma}(p,P) = \mc{C}\,{\Gamma}(-p,-P)^T\,\mc{C}^T$ with $\mc{C} = \gamma^4 \gamma^2$ and
$\mc{C}^T = \mc{C}^\dag = \mc{C}^{-1} = -\mc{C}$, cf. Appendix A of Ref.~\cite{Eichmann:2016yit}.

We determine the necessary input to Eq.~\eqref{onshell-current} from a combination
of DSEs and BSEs. The Bethe-Salpeter amplitude of a pseudoscalar meson
and the quark-photon vertex satisfy (in-)homogeneous BSEs
\begin{align}
   [\Gamma_M(p,P)]_{\alpha\beta} &= \int_q  [ \mathbf{K}(p,q,P)]_{\alpha\gamma;\delta\beta}   \nonumber \\
                                 & \quad \times [S_\q(q_+)\, \Gamma_M(q,P)\,S_\q(q_-)]_{\gamma\delta}\,,  \label{mesonBSE} \\
   [\Gamma^{\mu}_\q(p,P)]_{\alpha\beta} &= Z_2\,i\gamma^\mu_{\alpha\beta} + \int_q  [ \mathbf{K}(p,q,P)]_{\alpha\gamma;\delta\beta} \nonumber  \\
                                 & \quad \times [S_\q(q_+)\, \Gamma^\mu_\q(q,P)\,S_\q(q_-)]_{\gamma\delta}\,,  \label{qqgamma}
\end{align}
where $\mathbf{K}$ is the Bethe-Salpeter kernel, $Z_2$ the quark renormalization constant, $\q \in \{l,s\}$
and in both equations $q_\pm = q + \frac{\epsilon \pm 1}{2} P$. All results are independent on the momentum
partitioning parameter $\epsilon$ up to numerical artefacts. In a heavy-light system it turns out to be convenient
to adapt $\epsilon$ such that the complex quark momenta
tested by the BSE for both light and heavy quark are equally far away from the nearest non-analyticities of the respective quark
propagators defined by $\tilde{m}_q = \text{Im}\sqrt{p^2_\text{sing}}$. For the kaon this amounts to
$\epsilon = (\tilde{m}_l - \tilde{m}_s)/(\tilde{m}_l + \tilde{m}_s) \approx -0.15$.

The quark propagators $S_\q$ are given by their respective DSE,
\begin{equation}
\begin{split}
 S^{-1}_\q(p) &= Z_2\,(i\slashed{p} + Z_m m_q) \\
              &- Z_{1f} \,g^2 \,C_F\int_q i\gamma^\mu  \, S_\q(q) \, \Gamma^\nu_\text{qg}(q,p)\,D^{\mu \nu}(k) \,,
\end{split}
\end{equation}
where $m_q$ are the current-quark masses, $k=q-p$, $C_F=4/3$,
$D^{\mu \nu}$ is the dressed gluon propagator, $\Gamma^\nu_\text{qg}$ the dressed quark-gluon vertex
and $Z_2$, $Z_m$ and $Z_{1f}$ are renormalization constants.
The gluon propagator and quark-gluon vertex satisfy their own DSEs which include further $n$-point functions,
so that in all practical applications the tower of DSEs needs to be truncated.

In the following we work in Landau gauge and use the rainbow-ladder truncation. Together with
more advanced schemes this truncation has been reviewed recently in Ref.~\cite{Eichmann:2016yit}.
One defines an effective running coupling $\alpha(k^2)$ that incorporates all dressing effects
of the gluon propagator and the quark-gluon vertex. One then replaces in the quark DSE
\begin{equation}
   Z_{1f}\,g^2\,\Gamma^\nu_\text{qg}(q,p)\,D^{\mu \nu}(k) \; \to \; Z_2^2\,\frac{4\pi\alpha(k^2)}{k^2}\,T^{\mu\nu}_k\,i\gamma^\nu
\end{equation}
with transverse projector $T^{\mu \nu}_k = \delta^{\mu\nu} - k^\mu k^\nu/k^2$.
The BSE kernel $\mathbf{K}$ in Eqs.~(\ref{mesonBSE}--\ref{qqgamma}) is uniquely related to the
quark-self energy by the axialvector Ward-Takahashi identity. In rainbow-ladder truncation it is given by
\begin{equation}
   [ \mathbf{K}(p,q,P)]_{\alpha\gamma;\delta\beta} \; \to \; Z_2^2\,\frac{4\pi\alpha(k^2)}{k^2}\, i\gamma^\mu_{\alpha\gamma}\,T^{\mu\nu}_k\, i\gamma^\nu_{\delta\beta}\,.
\end{equation}
This construction satisfies chiral constraints such as the Gell-Mann--Oakes-Renner relation and ensures the
(pseudo-)Goldstone boson nature of the pseudoscalar mesons.

Once the effective interaction $\alpha(k^2)$ is specified, all elements of the calculation of the form
factors follow without additional adjustments. Similar to our previous work on the pion EMFF
\cite{Eichmann:2019tjk} we use the Maris--Tandy model, Eq.~(10) of Ref.~\cite{Maris:1999nt}, with a
convenient redefinition of their parameters $\{\omega,D\}$ to $\{\Lambda,\eta\}$ via $\omega D = \Lambda^3$ and $\omega=\Lambda/\eta$.
The scale $\Lambda=0.74$ GeV is fixed to reproduce the experimental pion decay constant $f_{\pi} = 92.4(2)$ MeV.
The variation of $\eta= 1.85 \pm 0.2$ changes the shape of the quark-gluon interaction at small momenta,
cf.~Fig. 3.13 in Ref.~\cite{Eichmann:2016yit}, and we use it as a rough estimate of the truncation error
similar to \cite{Eichmann:2019tjk}. In the DSE and BSE we work in the isospin symmetric limit of equal up/down
quark masses. With a current light quark mass of $m_q=3.57$ MeV at a renormalization point $\mu=19$~GeV we
obtain a pion mass of $m_{\pi^0} = 135.0(2)$ MeV. With the strange-quark mass fixed at $m_s=85$ MeV
we obtain a kaon mass of $m_{K} = 495.0(5)$ MeV.

\section{Results}

   \begin{figure*}[!t]
    \begin{center}
    \includegraphics[width=0.95\textwidth]{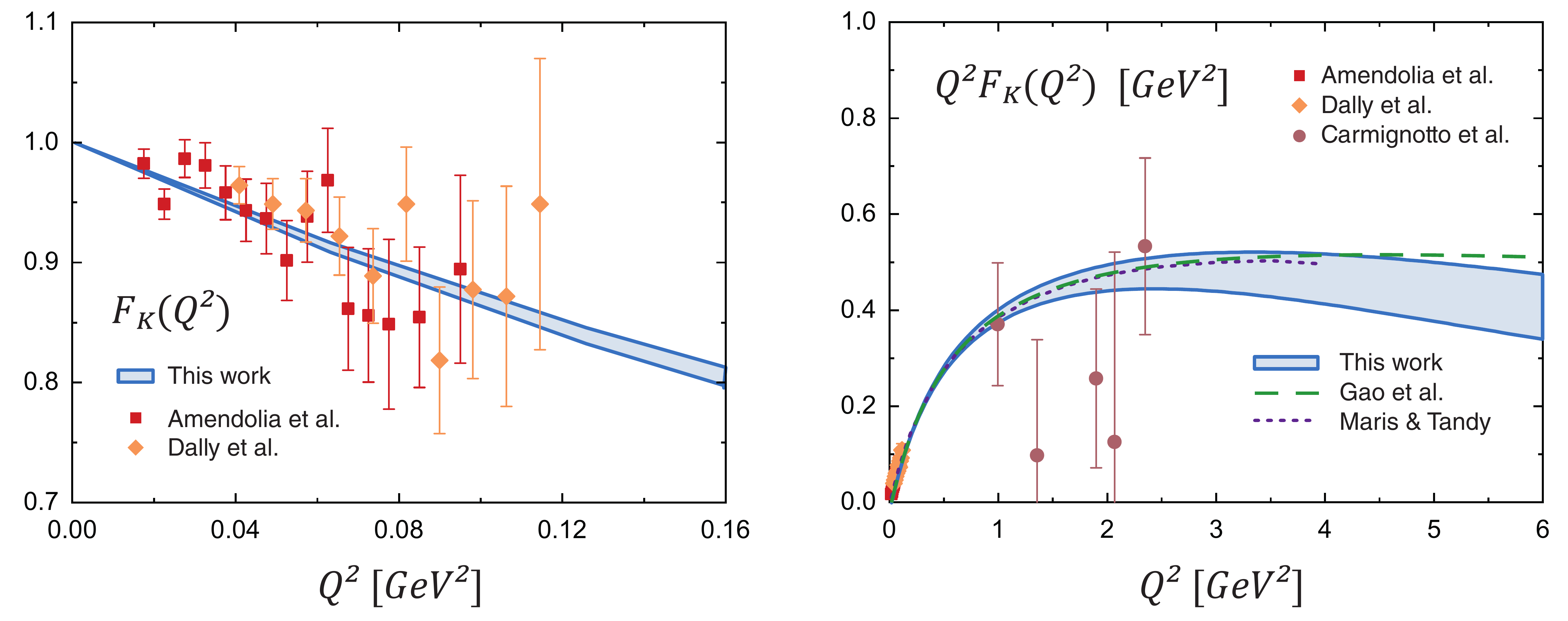}
    \caption{The kaon electromagnetic form factor as a function of the squared photon momentum.
    The experimental data have been extracted from
    Refs.~\cite{Dally:1980dj,Amendolia:1986ui,Carmignotto:2018uqj}.
    Previous results from DSEs have been obtained in \cite{Maris:2000sk,Gao:2017mmp}. \label{pionFF} }
    \end{center}
    \end{figure*}

\subsection{Kaon electromagnetic form factor}\label{sec:pion_ff}
    The EMFF of the kaon $F_{K^+}(Q^2)$ in the rainbow-ladder truncation described above has been
    determined previously in Refs.~\cite{Maris:2000sk,Gao:2017mmp}. Technical differences occur in the
    treatment of the quark-photon vertex: whereas in \cite{Maris:2000sk} only the leading five of eight
    tensor structures in the vertex have been taken into account, in \cite{Gao:2017mmp} the vertex has been
    represented by an ansatz in terms of quark dressing functions. Here (as in our previous work
    \cite{Eichmann:2017wil,Weil:2017knt,Eichmann:2019tjk}) we solve the BSE for the vertex including all
    tensor structures but neglecting the (weak) dependence of the dressing functions on the angle
    $k \cdot Q/(\sqrt{k^2} \sqrt{Q^2})$ between the photon momentum and the relative momenta of the quarks.
    In the momentum range relevant for $a_\mu$ all three approaches lead to very similar results
    rendering the technical differences marginal.

    In Fig.~\ref{pionFF} we show our results as a function of the squared photon momentum. The
    error band includes two types of systematic errors: (i) the variation of
    the parameter $\eta$ in the effective coupling and (ii) an estimate of the accumulating
    error due to neglecting the angular dependence in the quark-photon vertex. While
    the first error is straightforward to calculate, we estimated the second one by enforcing
    that the vertex should become bare if one quark momentum vanishes and the other becomes large,
    which introduces an angular dependence and leads to non-trivial relations between its dressing functions.
    Our results are compared to the experimental data extracted from
    \cite{Dally:1980dj,Amendolia:1986ui,Carmignotto:2018uqj}; the agreement is quite satisfactory,
    although the large error bars on the experimental data do not allow a stringent test
    of our approach. In the domain $Q^2 \sim 0 \dots 6$ GeV$^2$, the numerical results
    are well described by a monopole ansatz supplemented with an additional contribution which
    only becomes important at intermediate and large momenta,
    \begin{align}\label{pifit}
    F_{K^+}(Q^2) = \frac{1}{1 + Q^2/L^2} \frac{1 + c \, Q^4/L'^4}{1 + Q^4/L'^4}
    \end{align}
    with $c=0.1$ and the scales $L = 0.81(2)$ GeV and $L' = 2.9(3)$ GeV.
    The corresponding squared charge radius of the kaon is given by
    \begin{align}\label{kaoncharge}
    \langle r^2 \rangle_{K^+} = 0.36\, (2) \,\,\text{fm}^2\,.
    \end{align}    
    The calculation of the kaon EMFF beyond the $Q^2$ range displayed in Fig.~\ref{pionFF}
    faces a number of technical difficulties, which have been discussed in \cite{Eichmann:2019tjk}.
    For the large $Q^2$ contributions to the kaon-box in $a_\mu$ we extrapolate
    the fit outside the $Q^2 \sim 0 \dots 6$ GeV$^2$ domain.

    Compared to the pion EMFF, which follows a monopole behaviour for a large range of
    $Q^2$ values \cite{Eichmann:2019tjk}, the kaon EMFF deviates from a monopole already
    at comparably small momenta: first deviations become visible at about $Q^2=1$ GeV$^2$
    and sizeable from $Q^2=2$ GeV$^2$ on. These deviations have been discussed in
    detail in Ref.~\cite{Gao:2017mmp} and attributed to the differences in the distribution
    of light and strange quarks inside light mesons. Future high precision data at JLAB
    will be able to test this prediction from the DSE-approach.

    For the purpose of this work, the deviations from the monopole behavior are not very relevant. 
    The integrations involved in the calculation of the kaon-box diagram are strongly dominated 
    by momenta below $Q^2 \le 1$ GeV$^2$ such that the monopole scale $L$ in (\ref{pifit})
    is the most important quantity for $a_\mu$. We will quantify and discuss this issue below.

\subsection{Box-contributions to the anomalous magnetic moment of the muon}\label{sec:Results}
Before we study the kaon-box contribution to $a_\mu$ we report on an improved calculation of the
pion-box contribution as compared to the previous work \cite{Eichmann:2019tjk}. There the primary
focus was on the meson-exchange contributions to HLBL, which include the leading
contribution from neutral $\pi_0$-exchange. Since we are working in the isospin symmetric limit of QCD
all our pions have the same mass fixed by the experimental mass of the neutral pion, i.e.
$m_{\pi^0} = 135.0(2)$ MeV. This is also the mass that was used in \cite{Eichmann:2019tjk}
for the charged pion in the pion-box
diagram, leading to $a_\mu^{\pi^\pm-\text{box}} = -16.3 \,(2)(4)  \times 10^{-11}$, where the first error
was due to the variation of the model parameter $\eta=1.85 \pm 0.2$ and the second accounted for the numerical error.

The leading order effects of isospin breaking for that diagram is the mass difference
between the neutral and the charged pion $m_{\pi^\pm} = 139.57$ MeV \cite{Patrignani:2016xqp}.
Using the same pion form factor as in \cite{Eichmann:2019tjk} but taking into account the experimental mass
for the charged pion we obtain for the pion box contribution to HLBL
\begin{align}\label{pion-box}
a_\mu^{\pi^\pm-\text{box}}   &= -15.7 \,(2)(3)  \times 10^{-11}\,,
\end{align}
where again the first error is due to the variation of the model parameter and the second reflects the slightly
increased numerical precision as compared to \cite{Eichmann:2019tjk}. As a result of the adjustment of the charged
pion mass, the central value of our result moves closer to the result of the data-driven dispersive one
$a_\mu^{\pi^\pm-\text{box}} = -15.9 \,(2)  \times 10^{-11}$ \cite{Colangelo:2017fiz} and agrees very well within
error bars. Furthermore we now accurately reproduce their control-calculation with a VMD-type form factor; we obtain
$a_\mu^{\pi^\pm-\text{box-VMD}} = -16.4 \,(2)  \times 10^{-11}$. Thus we identified the pion mass as the
main source of the (slight) previous discrepancy between the value given in \cite{Eichmann:2019tjk}
and the dispersive approach. The
different numerical procedures (we use a nine-dimensional Monte-Carlo integration similar to
Ref.~\cite{Kinoshita:1984it}, whereas the authors of \cite{Colangelo:2017fiz} use algebraic methods to
simplify the numerical problem) seem not to affect the central value. The comments in \cite{Eichmann:2019tjk}
regarding the size of the numerical error of our calculation, however, remain valid.

Finally we present our results for the kaon-box contribution using the physical charged kaon mass
 $m_{K^\pm} = 493.68$ MeV \cite{Patrignani:2016xqp}. We obtain
\begin{align}\label{kaon-box}
a_\mu^{K^\pm-\text{box}}   &= -0.48 \,(2)(4)  \times 10^{-11}\,,
\end{align}
where the first error is due to the variation of the model parameter and the second accounts for the
numerical error of our Monte-Carlo integration. We find that the contribution due to the kaon box is
only about 3 \% of the value of the pion-box diagram and therefore truly subleading. This is in
qualitative agreement with previous determinations of the kaon box using form factors from vector
meson dominance (VMD) considerations \cite{Bijnens:1995xf} or hidden local
symmetry (HLS) \cite{Hayakawa:1995ps,Hayakawa:1996ki}.

In order to quantify the impact of the deviations from the pure monopole at large $Q^2$, i.e. $c \ne 1$ in
Eq.~(\ref{pifit}), we also determined the pure monopole contribution, $c=1$, and obtained: 
$a_\mu^{K^\pm-\text{box-pure monopole}} = -0.49 \,(2)(4)  \times 10^{-11}$. As expected, the impact is
very small. 

We also compare to a control calculation using a VMD-type form factor given by
\begin{align}
F_{K^+}^{VMD}(Q^2) = 1-\frac{Q^2}{2}&\left(            \frac{1}{M_\rho^2  +Q^2}\right.
                                           +\frac{1}{3}\frac{1}{M_\omega^2+Q^2}
 \nonumber\\
                                         &\,\,\left.
                                         +\frac{2}{3}\frac{1}{M_\phi^2  +Q^2}\right)\,.
\end{align}
This form factor is slightly larger than our result from DSEs/BSEs and consequently
leads to a slightly larger absolute value for the contribution to $a_\mu$:
$a_\mu^{K^\pm-\text{box-VMD}}  = -0.54 (4)  \times 10^{-11}$. The corresponding calculation 
using the very same VMD-type form factor but the integration method of \cite{Colangelo:2017fiz} 
leads to $a_\mu^{K^\pm-\text{box-VMD}}  = -0.50  \times 10^{-11}$ \cite{Stoffer:2019}. Thus although
our central value is a little larger we find agreement within error bars.

Finally, we wish to note that besides the charged meson loop contributions, there is also a contribution
from the neutral kaon. This is because the corresponding form factor is only zero at vanishing momentum
but has non-trivial momentum dependence at finite space-like momenta, see e.g. \cite{Maris:2000sk}. 
Given that this contribution is certainly highly suppressed, we do not attempt here to determine its size.

\section{Summary}
In this work we have presented a calculation of the charged kaon-box contributions to hadronic light-by-light
scattering based on a functional approach to QCD via Dyson-Schwinger and Bethe-Salpeter equations. We employed
the same rainbow-ladder truncation for the quark-gluon interaction as in our previous work on the
pseudoscalar meson pole contributions to HLBL, 
$a_\mu^{\text{PS-pole}}   = 91.6 \,(1.9)  \times 10^{-11}$ \cite{Eichmann:2019tjk}. 
We updated and improved our previous result
on the pion-box contribution by taking into account isospin breaking effects in the pion masses.
Our result, Eq.~(\ref{pion-box}), is in excellent agreement with the most recent dispersive result 
of Ref.~\cite{Colangelo:2017fiz}.
Based on this agreement we consider our result for the kaon-box contribution, Eq.~(\ref{kaon-box}), 
quantitatively meaningful.
Its value is only about 4 \% of the one for the pion box and thus only gives a very small contribution to $a_\mu$.
Note again that the error bar given in our final result does not contain all systematic errors:
there is an additional error due to truncation effects, which is very hard to quantify and consequently
has been left out. Since the kaon-box contribution, however, is very small we expect this omission to have only
a very marginal if not negligible effect on the total error of HLBL.

\vspace*{3mm}\noindent{\bf Acknowledgments}\\
We are grateful to Johan Bijnens, Gilberto Colangelo, Bastian Kubis, Massimiliano Procura 
and Peter Stoffer for inspiring discussions and critical comments.
We give special thanks to Esther Weil for discussions and a critical reading of the manuscript.\\
The idea for this work has been born in a discussion session of the workshop ``Hadronic contributions to (g-2)$_\mu$'' 
(INT 19-74W) at the University of Washington, 09/09-09/13 2019. CSF is grateful for financial support 
by the INT during the workshop.\\
This work was furthermore supported by the Helmholtz International Center for FAIR within
the LOEWE program of the State of Hesse, by the Helmholtz centre GSI in Darmstadt, Germany
and by the FCT Investigator Grant IF/00898/2015.

\bibliography{paper}

\end{document}